\documentclass[a4paper,11pt]{article}
\usepackage{jinstpub} 
\usepackage{lineno}
\usepackage{multirow}
\usepackage{graphicx}
\usepackage{lipsum}
\usepackage{natbib}

\usepackage[inkscapelatex=false]{svg}

\title{\boldmath The design of the data acquisition system for SPD experiment}

\author[a]{L. Afanasyev}
\author[b]{A. Berngardt}
\author[a]{A. Boikov}
\author[b]{V. Borshch}
\author[c]{A. Boykov}
\author[c]{D. Cherepkov}
\author[b]{D. Erofeev}
\author[a]{V. Frolov}
\author[a]{K. Gritsay}
\author[a]{A. Isupov} 
\author[c]{K. Kotov}
\author[c]{D. Kozyrev}
\author[a]{A. Kulikov}
\author[c]{O. Mamoutova}
\author[c]{N. Reshetikhin}
\author[c]{D. Ryabikov}
\author[a]{S. Tereshchenko}
\author[a]{V. Tereshchenko}

\affiliation[a]{Joint Institute for Nuclear Research, Dubna, Russia}
\affiliation[b]{The National Research Tomsk State University, Tomsk, Russia}
\affiliation[c]{Peter the Great St.Petersburg Polytechnic University, St. Petersburg, Russia}

\emailAdd{boikov@jinr.ru}
\emailAdd{aleksandr.boikov@cern.ch}

\abstract{The Spin Physics Detector (SPD) experiment at the NICA collider in JINR aims to investigate the spin structure of nucleons and spin-related phenomena. The combination of the number of background processes, the event rate and conditions for event selection makes the use of a classical trigger system impractical, requiring a triggerless data acquisition (DAQ) system. The DAQ system is designed to ensure precise time synchronization, efficient data collection, and high-throughput processing. Its architecture combines commercially available FPGA-based modules and high-speed optical interfaces with custom-developed components based on widely accessible technologies.  This approach provides scalability from 180,000 at the initial stage of the experiment to more than 600,000 detector channels in the final configuration and supports data rates up to 20 GB/s or more. The modular system design ensures adaptability for future upgrades while maintaining high efficiency and reliability. Such an approach makes the DAQ system suitable for other high-rate nuclear physics experiments.}


\begin{document}
\maketitle
\flushbottom

\section{Introduction}

The main goal of the SPD (Spin Physics Detector) experiment placed at the Nuclotron based Ion Collider fAcility (NICA) at JINR (Dubna) is a study the spin structure of the proton and deuteron and the other spin-related phenomena with polarized proton and deuteron beams at a collision energy up to $\sqrt{s}=27$~GeV \citep{TDR}.

NICA accelerator complex consists of two linear accelerators, transport beam lines, two synchrotrons and two storage rings of the collider. NICA has two intersection points to observe interactions between particles of the two beams \citep{NICA}.

The SPD experimental setup is designed as a universal $4\pi$ detector with advanced tracking and particle identification capabilities. SPD will be realized in two stages. The following detectors are involved in the first stage: 
\begin{itemize}
\item Micromegas-based Central Tracker (MCT)
\item Straw Tracker (ST)
\item Range System (RS)
\item Beam-Beam Counters (BBC)
\item Zero Degree Calorimeters (ZDC)
\end{itemize}

In the second phase, additional subsystems are going to be implemented, primarily for particle identification:

\begin{itemize}
\item Silicon Vertex Detector (MAPS or DSSD) instead of MCT
\item Time-Of-Flight system (TOF)
\item Aerogel Cherenkov counters (FARICH)
\item Electromagnetic Calorimeter (E-Cal)
\end{itemize}

At the maximum center mass energy $\sqrt{s} = 27$~GeV and maximum luminosity $10^{32}$~$\;\mathrm{cm}^{-2}\mathrm{s}^{-1}$ the interaction rate was estimated at $3 \times 10^6$~$\;\mathrm{s}^{-1}$. The designed bunch crossing period is 76 ns.

The combination of the event rate, proximity of background processes and signal, and the selection of physics signals — which rely on momentum and vertex reconstruction will require a hardly realizing classical hardware-based trigger system. Because of this, a free-running (triggerless) data acquisition system is proposed. This system will enable the recording of all collisions without preliminary selection, ensuring a greater flexibility in subsequent analysis. This approach is particularly crucial for investigating rare processes, as data on such events might be lost when using a conventional trigger system.

In our approach all signals exceeding the thresholds in the front-end electronics (FEE) are read out with their timestamp. The DAQ system organizes data into time-based units called slices (10--100~\textmu s) and frames (0.1--10~s). Each slice is uniquely identified by a frame number, indicating its position within the data taking period (Run), and a slice number, denoting its order within the frame. 

A global clock, transmitted from the main reference source, is distributed using a network based on White Rabbit (WR) technology~\citep{WR_main}. Each element of the system is synchronized to this clock with a frequency of 125 MHz (8 ns period) and relative jitter of $\leq 50$~ps.

Two sets of commands are used to control the DAQ system: synchronous with the global clock and asynchronous. Synchronous commands provide the electronics with signals to start/stop a slice and start/stop a frame. Asynchronous commands provide initialization of the electronics and the procedure for starting the run.

Timestamps for each hit are measured relative to the start of the frame, ensuring time correlation only within the same frame. Consequently, a frame defines the maximum time interval for detecting time-correlated events in the SPD detector. This structure ensures precise event timing while maintaining data integrity and synchronization across the system. The DAQ system collects data from the front-end electronics, forms them into data files of a specific format, and writes these data files to a fast intermediate storage \citep{TDR} .

In addition, DAQ will be able to provide configuration of FEE, monitoring status of FEE, and possibility to transfer a new firmware to FEE.

\section{Estimation amount of data}

In the first and second phases, different types of detectors and numbers of channels will be used. Summary information about the detector channels is given in table~\ref{tab:daq-ch}.  For correct data collection the FEE have to meet the requirements of a free-running DAQ, such as self-triggered operation, on-board digitization, and the inclusion of timestamps in the output data format.

With certain approximations, the total data flux was estimated to be approximately 20 GByte/s. This estimation was based on a combination of simulations and experimental results from beam tests of detectors from other experiments (PANDA, MPD, ALICE), along with parameters of the relevant front-end electronics, either existing or under development. The data flux was estimated at the event rate of $3 \times 10^6$~s$^{-1}$ and multiplicity value equals to 8 for charged particles and 6 for neutral ones per event. This result is providing a rough but reasonable value for the expected data throughput. In the first stage of experiment the data flux must be smaller by up to two orders of magnitude. A total number of channels at the first stage does not exceed 180 thousand. For the full scale this number is expected to be about 600 thousand.

\sloppy
\begin{table}[ht]
\begin{center}
\begin{tabular}{|c|c|c|c|c|}
\hline
\small
\textbf{Sub-detector} & \textbf{Info} & \textbf{Ch} & \textbf{Ch/FEE} & \textbf{Outputs} \\ \hline
\multicolumn{5}{|c|}{\textbf{PHASE 1}} \\ \hline
Micromegas            & T + A & 5384                & 128      & 43  \\ \hline
Straw tracker         & T + A & 30512               & 64      & 478 \\ \hline
BBC                   & T + A & 1056                & 64       & 18  \\ \hline
BBC MCP               & T     & 64                  & 32       & 2   \\ \hline
Range system          & T     & 137600              & 192      & 717 \\ \hline
ZDC                   & T + A & 2100                & 64       & 34  \\ \hline
\textbf{Total (max)}  &       & \textbf{176716}     &          & \textbf{1292} \\ \hline
\multicolumn{5}{|c|}{\textbf{PHASE 2}} \\ \hline
Vertex DSSD           & A     & 327680              & 640      & 512 \\ \hline
Vertex MAPS           & A     & 12000               & 8        & 1500 \\ \hline
Straw tracker         & T + A & 30512               & 64      & 478 \\ \hline
Calorimeter           & T + A & 27168               & 64       & 425 \\ \hline
TOF                   & T     & 12228               & 8        & 1529 \\ \hline
FARICH                & T     & 70144               & 64       & 1096 \\ \hline
BBC                   & T + A & 1056                & 64       & 18  \\ \hline
BBC MCP               & T     & 64                  & 32       & 2   \\ \hline
Range system          & T     & 137600              & 192      & 717 \\ \hline
ZDC                   & T + A & 2100                & 64       & 34  \\ \hline
\textbf{Total (DSSD)} &       & 608552              &          & 4811 \\ \hline
\textbf{Total (MAPS)} &       & \textbf{283076}     &          & \textbf{5800} \\ \hline
\end{tabular}
\caption{Number of detector outputs to DAQ, where T stands for the mean time and A for the amplitude (or charge).}
\label{tab:daq-ch}
\end{center}
\end{table}
\section{Structure}

Logically, the DAQ system can be divided into following subsystems:
\begin{itemize}
\item Time synchronization system (TSS)
\item Readout chains
\item Slice building system
\end{itemize}

\begin{figure}
    \centering 
    \includegraphics[width=1\textwidth]{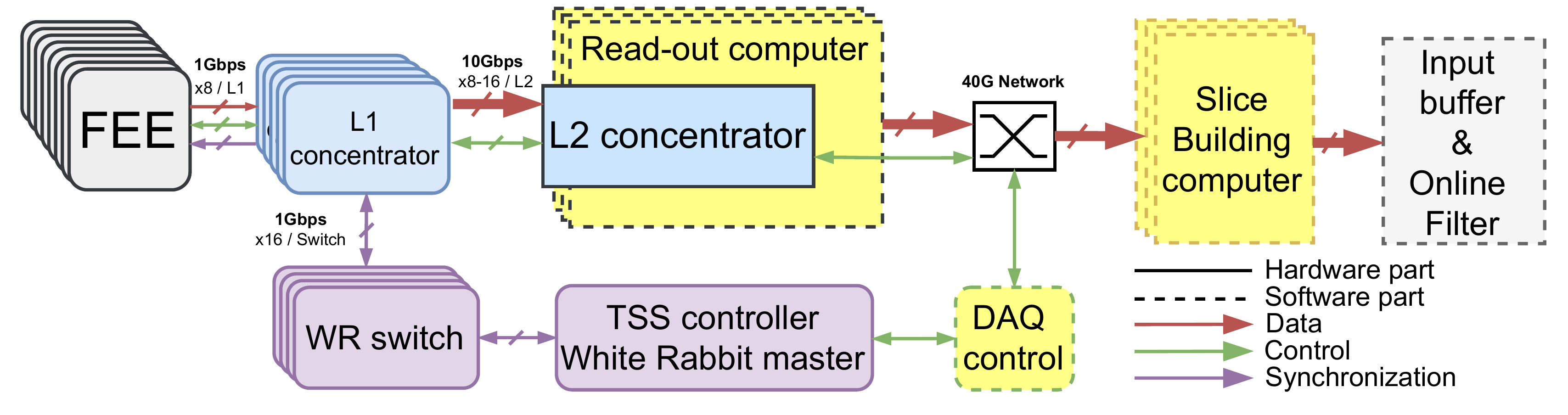}
    \caption{DAQ structure.} 
    \label{fig_struct}
\end{figure}

The time synchronization system serves for distributing the global clock signal and synchronous commands throughout the SPD setup. Data from all subdetectors transmit through two levels of data concentrators (L1 and L2), called readout chains, to the set of readout computers. There will be about 150 readout chains at the full scale setup. Each readout chain transmits data from their part of subdetectors to the separate readout computers. As a result, the data from one slice is distributed across all the readout computers. The purpose of slice building systems is to collect data from all readout computers, build a complete slice and write it to a fast intermediate storage. 
Structure of the DAQ system is shown in figure~\ref{fig_struct}.

All information regarding DAQ configuration, including parameters such as slice and frame length, enabled subdetectors, the configuration of front-end electronics (e.g., calibration settings and channel mapping), as well as experimental conditions (e.g., beam energy and polarization), will be recorded and stored in a dedicated database.

\section{Time Synchronization System}

The Time Synchronization System ensures precise data timestamping across the distributed SPD DAQ system. The TSS generates the global clock for each FEE as well as synchronous commands to define data unit time intervals with sufficient accuracy for subsequent event reconstruction. 

As the deviations of oscillator frequencies between distributed devices cannot be avoided, the task of coherent event detection and detectors' data collection is crucial for the quality of data in distributed experimental setups. Proper clock and time synchronization between devices provides a common time base for all of the data channels in the DAQ network, enabling reference information for the sampled data fragments that allows for proper data alignment during further processing of the data from distributed sources.  Examples of existing experimental physics setups typically include custom dedicated synchronization systems: dedicated optical fibers in a TTC at the LHC~\cite{varela_timing_2000} or deterministic latency messaging in a DAQ for CBM~\cite{lemke_unified_2010}.

Among existing synchronization technologies, IEEE1588-2019 is a standard for accurate time synchronization, and its high-precision profile (White Rabbit) is used to achieve subnanosecond accuracy. White Rabbit (WR) technology employs a hierarchical network architecture over fiber, allowing for relatively easy scaling of the network and information exchange over the same media the synchronization is supported on. WR nodes perform hardware clock recovery and propagation for syntonization, include digital implementation of the Didgital Dual Mixer Time Difference phase detector for precise time interval measurements, and extend the PTPv2 protocol to support the establishment of a WR link ~\cite{daniluk_white_2011}. As a result, the local clock of each device is automatically synchronized with the global clock (at 125/62.5 MHz) with an achievable accuracy of better than 1 ns. Open-source gate-level and firmware implementation of the WR PTP Core allows for a wide variety of FPGA-based WR-capable devices. 

The White Rabbit enables the synchronization on the connected devices which is achieved after performing the necessary calibration procedure. The procedure yields a series of calibration coefficients which mitigate the propagation delay of the global clock throughout the network and are device-dependent. The SPD will be built in two stages. The second stage of the experiment requires synchronization of 5 times more L1 concentrators than in the first stage. Standard network connections will simplify the upgrade procedures and enable the reusability of the calibration coefficients. To support the stepwise integration of L1 concentrators into the setup, the WR calibration values are stored externally and are remotely imported onto the WR devices in the network. 

SPD requires that the global clock at each DAQ element has a relative phase deviation of no more than one nanosecond and a phase jitter of no more than 50 ps rms. Since different SPD subdetectors may have more strict or relaxed requirements, it is important to carefully assess the limits of the WR accuracy and precision for each subdetector. In the TSS network, the WRS-18A devices from Sync Technology are going to be used as 18-port network switches, and the CuteWR-A7 devices, also from Sync Technology, are used as early prototypes of the TSS nodes. Experimental evaluation of the WR devices has shown their acceptable baseline jitter and clock skew performance for SPD DAQ. Detailed accuracy and precision analysis, including jitter distribution and temperature effects on WR devices, is reported in ~\cite{kozyrev_WR_2024}.

\begin{figure}
    \centering 
    \includegraphics[width=1\textwidth]{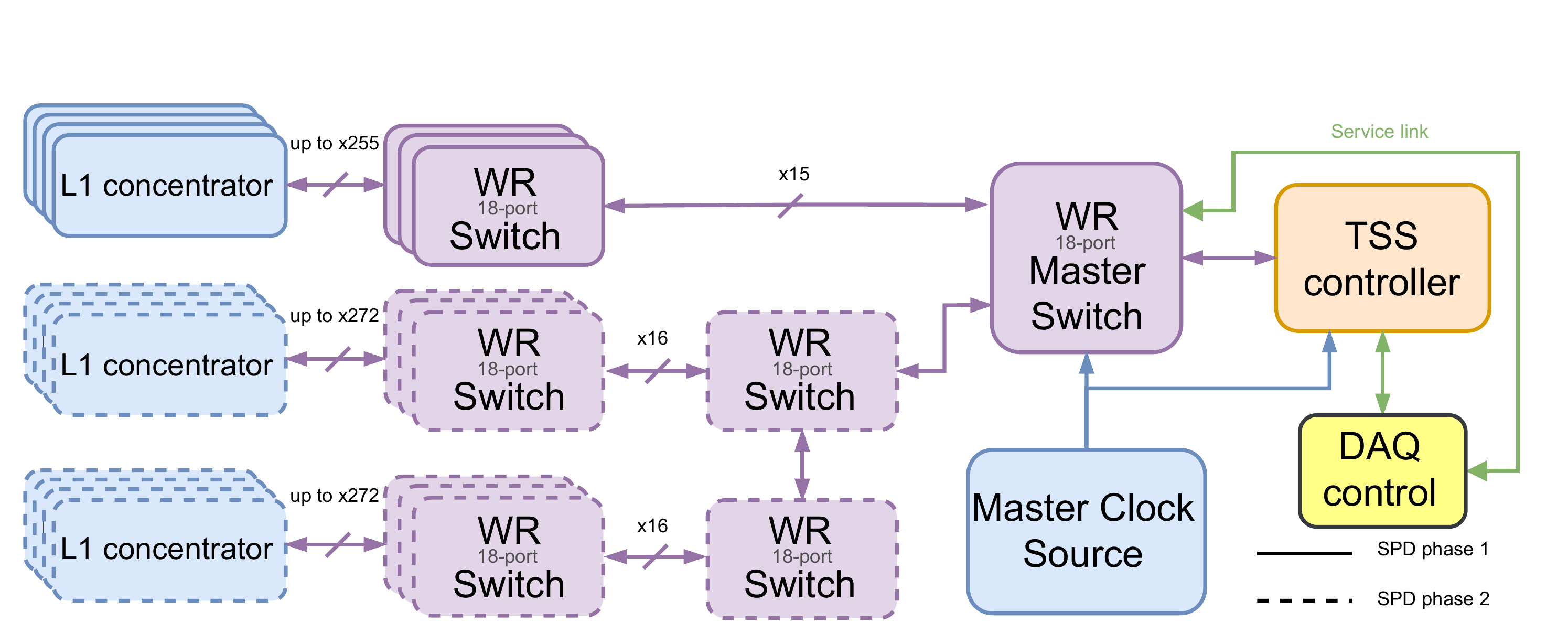}
    \caption{TSS network topology.} 
    \label{DAQ_TSS_Topology}%
\end{figure}

The TSS uses a WR-capable Ethernet network as the synchronization medium for all elements of the TSS: the TSS controller and the TSS nodes ~\cite{ryabikov_WR_2024}. The TSS controller creates and broadcasts over the network a schedule of the synchronous commands (see below~\hyperref[sec:commands]{DAQ commands}) based on the run parameters from the DAQ run control. WR-capable TSS nodes are integrated into L1 concentrators and execute the broadcasted schedule by generating synchronous commands. The Master Clock Source will be the stable global clock and PPS source for the WR network synchronization. All L1 concentrators and the TSS controller serve as WR slaves continuously aligning their local oscillators to the global clock for stable synchronization. Only one port of the switches is a WR slave to the upper-level switch providing other ports to other devices to reference to. The master switch receives the clock from the dedicated SMA port, freeing one additional SFP port to be a WR master. The topology of the synchronization system is shown in figure~\ref{DAQ_TSS_Topology}.


The synchronous commands are transmitted to the FEE in the form of pulse trains accompanied by the globally synchronized clock signal. The synchronous commands determine the time borders of the frames and slices inside the frames. The accuracy of the global clock signal at the FEE inputs determines the consistency of timestamping throughout the system.
\section{DAQ commands}
\label{sec:commands}


\begin{figure}
    \centering 
    \includegraphics[width=1\textwidth, keepaspectratio]{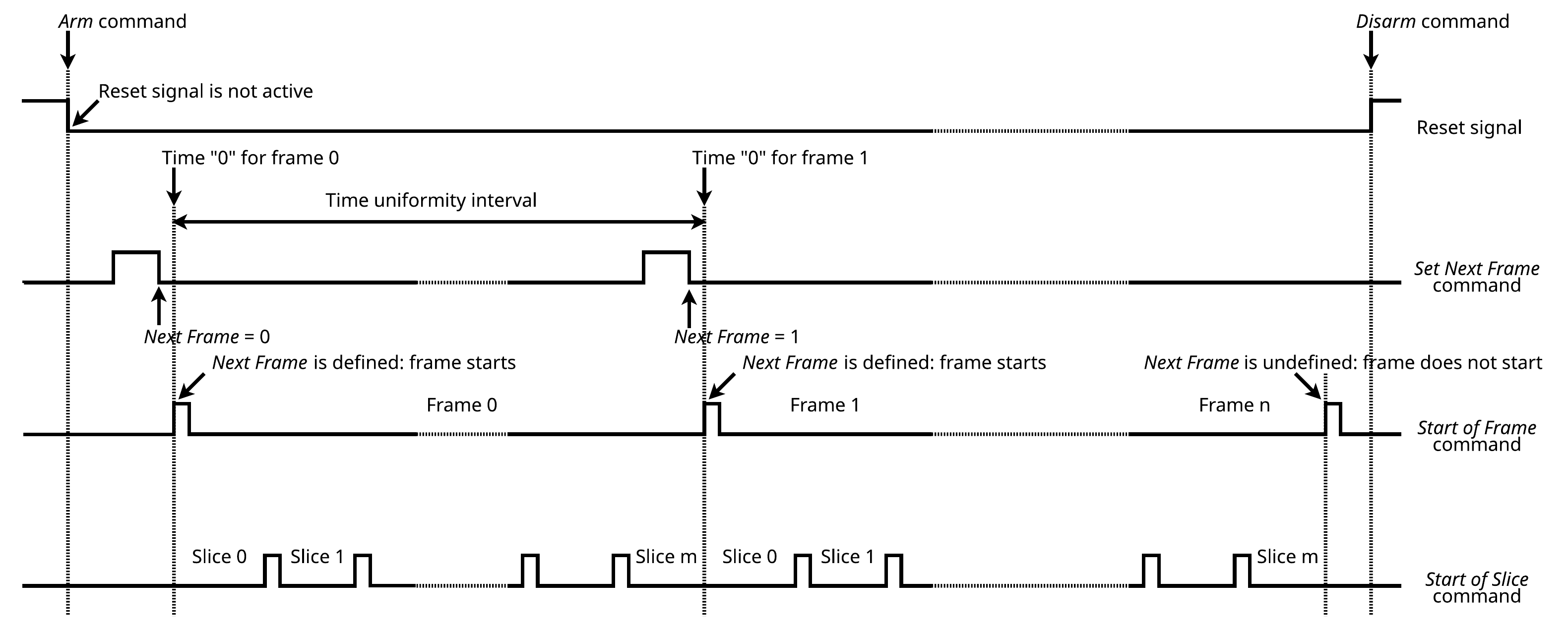}
    \caption{The time structure of the run (simplified).}
    \label{daq-run_start}
\end{figure}

The control of the readout chain is facilitated through two types of commands: synchronous commands aligned with the global clock and asynchronous commands \citep{TDR}.

Synchronous commands, transmitted along with the global clock, are broadcasted to the L1 concentrators via the TSS. The primary synchronous commands include:

{\bf Set Next Frame:}
This command prepares the system for the next frame by transmitting its number. It ensures atomicity, meaning the FEE fully receives number of next frame.

{\bf Start of Frame:}
This command completes the current frame (if active) and initiates a new frame using the preloaded frame number. Simultaneously, it starts the first slice of the new frame. At the end of the frame, the last slice is automatically stopped.

{\bf Start of Slice:}
This command concludes the current slice and starts a new one within the same frame. The slices are automatically numbered by the FEE, with the slice numbering reset to zero by the Start of Frame command.

An example of the work is shown in figure~\ref{daq-run_start}.

Asynchronous commands are transmitted to the L1 concentrators by the L2 concentrators. These commands can target individual modules or all modules connected to a specific L1 concentrator. Standard asynchronous commands include:

{\bf Disarm:}
Activates the reset signal line from the L1 concentrator to the FEE. In this state, the FEE ignores all synchronous commands and initiates its reset process.

{\bf Arm:}
Deactivates the reset signal line, enabling the FEE to process synchronous commands.

In addition to these standard commands, each FEE may support a unique set of asynchronous commands for initialization and monitoring. Monitoring commands, which do not alter the FEE state, can be executed concurrently with synchronous commands. Asynchronous commands are transmitted to the FEE as addressed I$^2$C commands or over packet based dedicated line, ensuring precise module-specific control.

The combination of synchronous and asynchronous commands provides the necessary functionality to ensure precise control and stable operation of the data acquisition system.
\section{Readout chain}
\begin{figure}
    \centering 
    \includegraphics[width=1\textwidth, keepaspectratio]{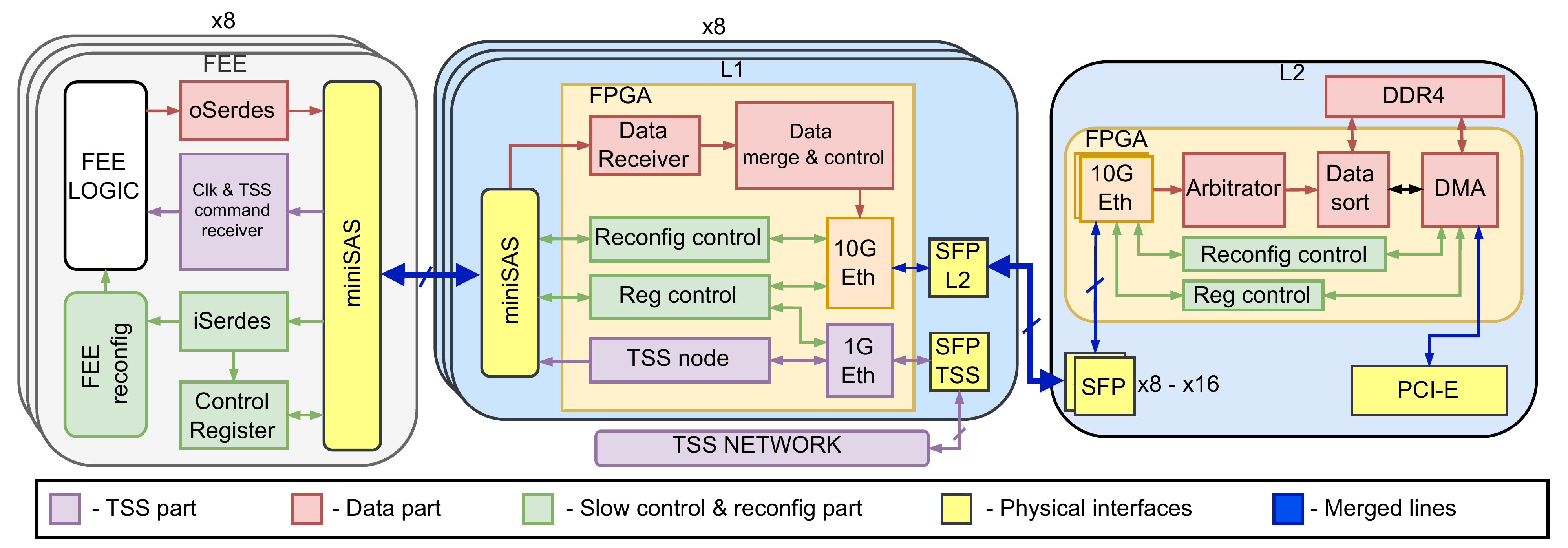}
    \caption{Readout chain.} 
    \label{readout_chain}
\end{figure}

The readout chain includes the FEE, a first-level data concentrator (L1 concentrator), and a second-level data concentrator (L2 concentrator). The L2 concentrator is installed in the readout computer. The general view of the readout chain is shown in figure~\ref{readout_chain}.

The FEE cards are linked to the L1 concentrator via copper flexible cables with MiniSAS (SSF-8087) connectors. The cable length can be up to 7 meters, depending on the position of the detector. MiniSAS cables are used for transmitting LVDS signals, including data transfer, the global clock, and signals for TSS commands. In addition to the LVDS signals, the cable also carries unipolar signals for slow control bus, reset, and presence detection. The slow interface is used to transmit parameters to the FEE and send special asynchronous commands. These commands are designed for controlling and monitoring the FEE and are specific for different electronics. 

The L1 concentrator board is designed to support the connection of 8 FEE, with a data transfer rate up to 1 Gbps per channel. The L1 concentrator will be based on FPGA chips, such as the Cyclone 10GX. Configuration and management data for the FEE cards are transmitted from the L2 concentrator via an optical link. The connection between L1 and L2 is expected to use a high-speed bidirectional optical channel with data rates up to 10 Gbps. Communication between concentrators is organized by using the Ethernet II (802.3 "Raw") protocol \cite{IEEE802.3}. 

The global clock signal and synchronous commands are generated and distributed by the TSS and transmitted to the L1 concentrators via a dedicated optical interface. Within each L1 concentrator, an integrated White Rabbit node manages the generation of commands for the FEE and the configuration of the White Rabbit network. The readout chain operates under the control of the TSS, ensuring that frames and slices begin and end according to the commands issued by the TSS. While data transmission is triggered by the front-end electronics in response to TSS commands, there is no strict time correlation between the issuance of these commands and the initiation of data transmission. 

At the L1 concentrator the status and data rate of each FEE will be monitored. This information can be obtained on request or in the periodic-message mode. In case of emergency situations, L1 will inform the DAQ control system. It is possible to transmit monitoring messages via a connection to the L2 concentrator or the WR network.

The L2 concentrator is implemented using a commercially available evaluation board with the PCIe interface (ALINX Z19-P with Zynq UltraScale+ MPSoC on board), installed in the readout computer. A single L2 concentrator can connect from 8 to 16 L1 concentrators, corresponding to a total from 64 up to 128 FEE cards. The firmware developed for these cards is responsible for receiving data from the L1 concentrators, pre-sorting and transmitting it to the readout computer. Data is transferred directly to the computer's memory (RAM) via the PCIe 3.0 x16 bus at a speed of 8–12 GBps. Additionally, the firmware manages the transmission and reception of control data to and from the L1 concentrators for the management of both the L1 concentrators and the FEE \citep{l2_concentrator_2025}.

Up to 4 L2 concentrator cards can be installed in a single readout computer, allowing for scalable data acquisition. This depends on the specific computer, the number of available PCIe 3.0 x16 slots and data rate for each L2 concentrator.

\section{Readout chain data format}
In our data acquisition system, the fundamental unit of information is a slice. It must include all the essential data required for the unambiguous reconstruction of all hits that occurred during one slice.

\begin{figure}
    \centering 
    \includegraphics[width=1\textwidth]{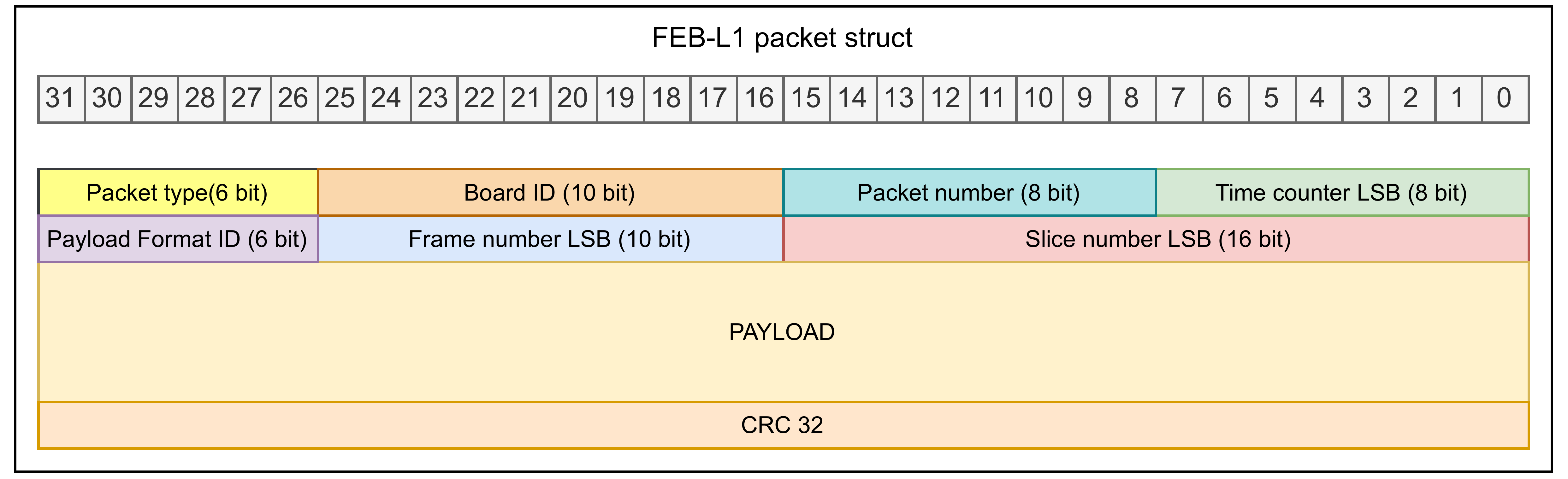}
    \qquad
    \includegraphics[width=1\textwidth]{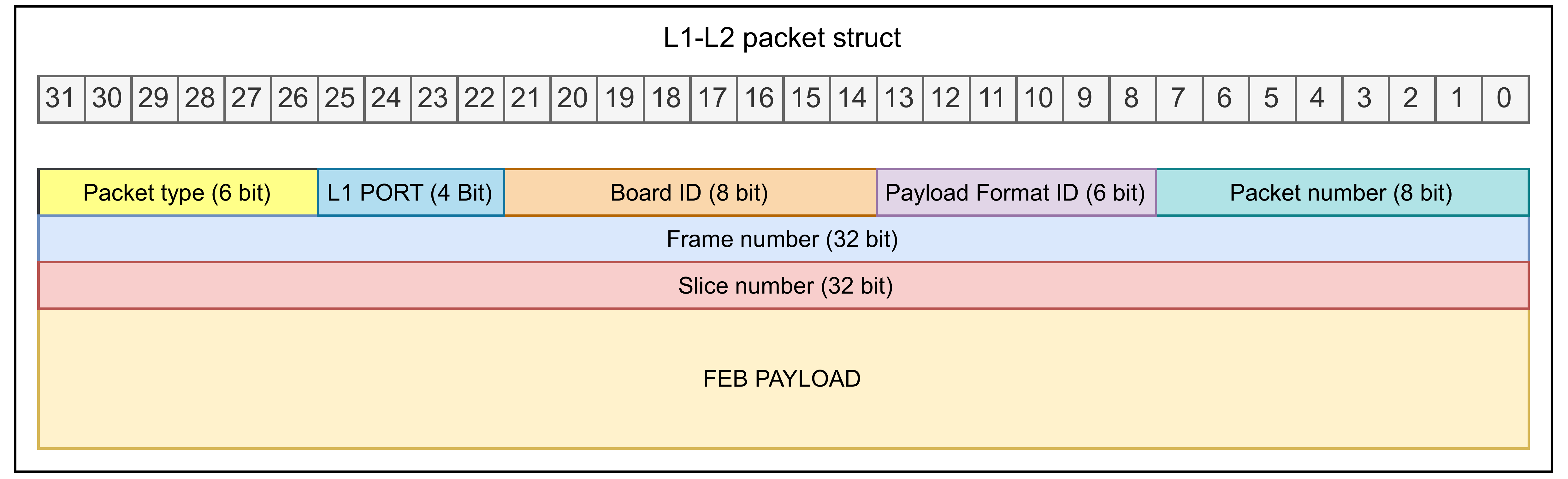}
    \qquad
    \caption{FEE-L1-L2 data format (preliminary).} 
    \label{daq-febdata}
\end{figure}

The data format is a critical determinant of system performance and efficiency. It must encapsulate the maximum amount of information while minimizing unnecessary overhead, as even a single additional byte per slice can increase the total data flow by 0.5 GB/s. To support efficient processing by computational systems, the data have to be aligned to a 32-bit word. If data elements from the front-end electronics deviate from this alignment, the components of the readout chain must ensure proper realignment. This consideration is particularly significant for DAQ system in the SPD detector, where precise data alignment is essential for maintaining high throughput and overall system performance. 
At the same time, the data (FEE payload) will contain all available information about hits: the hit location, the hit registration time, the amplitude of the measured signal, etc.


Each FEE has its own internal data structure describing the hits. However, each data block from the FEE includes a header with an ID for the data type, the LSB parts of the slice and frame numbers, the FEE ID, and a checksum. At the L1 level, the remaining bits of the frame and slice numbers, as well as the L1 port number, are added to this data. Preliminary data format for readout chain is shown in figure~\ref{daq-febdata}.

\section{Slice building system}
The data in the slice building system is processed in parts, each part covers a certain period of time. The natural choice for this time period is a frame. In some cases, the amount of data in the frame may be too large for convenient data processing. Therefore, the frame is programmatically divided into parts called chunks. A reasonable chunk length is about 1~s and can be chosen taking into account the actual data flow. Splitting the frame into chunks is performed on the border of the slices, if the frame length exceeds the specified length of the chunk. Thus, the chunk is a data processing unit in the slice building system. The chunk is invisible to the rest of the software.

Since different data structures of the slice building system can simultaneously contain slices related to different runs, 3 parameters must be used to address the slice: the run number, the frame number and the slice number. This imposes the condition that the run number must be unique. At the same time, the need to sort slices by time requires a monotonous increase in this number. In order to leave some freedom in choosing the numbering of runs, for the purposes of identifying the run, the run ID will be used --- an automatically generated monotonically increasing number that is assigned to each run. The run number will be used in situations where human interaction is expected.

Two numbers will be used to address the chunk: the run ID and the chunk number. The chunk number is obtained from the frame number and the slice number of the first slice belonging to the chunk.

\begin{figure}
    \centering 
    \includegraphics[width=0.8\textwidth, keepaspectratio]{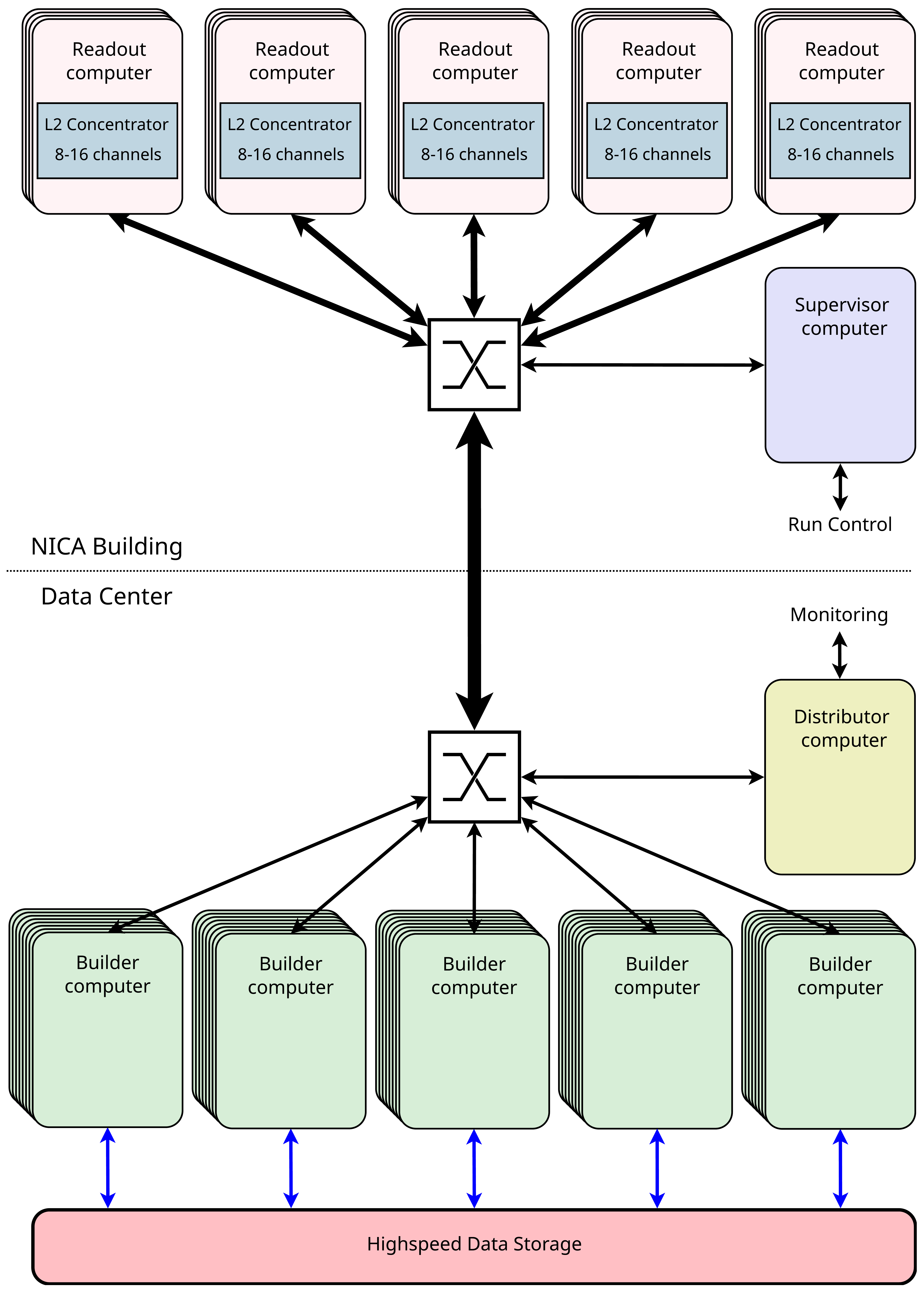}
    \caption{Slice building system.} 
    \label{fig:slice_build}
\end{figure}

The general view of the slice building system is shown in figure~\ref{fig:slice_build}. The slice building system consists of readout computers, a supervisor computer, builder computers, distributor computers and network switches. The various components of the system are distributed across 2 halls: SPD hall at NICA and NICA data center.

From the software point of view, the slice building system includes the following main processes: reader processes, supervisor process, builder processes and distributor processes.

The reader processes are performed on the readout computers, one process for each L2 concentrator. The main tasks of the reader process are: receiving data from the L2 concentrator, buffering it in RAM, and sending data to the builder processes according to their requests.

The reader process is controlled by the supervisor process. The main commands from the supervisor are to start/stop the run, request to send metadata, and drop a chunk. Data coming from the L2 concentrator, organized into slices, is stored in the data buffer. The buffering depth will range from tens of seconds to 1--2 minutes, depending on the available memory and data flow. For each chunk, the reader process fills in metadata, the main of which are the run ID, the chunk number, the number of slices read from the L2 concentrator, the number of slices sent to the builder process, and the timestamps of the beginning and end of reading the chunk from L2, the beginning and end of chunk transmission to the builder process. The metadata is passed to the supervisor process upon request from it. The request from the builder process to send data contains the run ID and the chunk number. The reader process sends the chunk data to the builder process slice by slice, releases the transferred slices from the data buffer, and updates the metadata structures.

The supervisor process runs on the dedicated computer. The system uses only one supervisor process, which works as a dispatcher for the entire slice building system. The supervisor process periodically reads metadata from all reader processes. Based on the information received, it makes a decision about the chunks whose reading from the L2 concentrators has been completed or stopped due to a timeout. A timeout situation is registered when data from any reader process did not appear during the specified time after the first data about this chunk from any other reader process has appeared. Incomplete chunks (with missing slices or with a completely missing chunk data from one or more reader processes) are deleted. Complete chunks are placed in a  queue of chunks ready for building (ready queue).  A chunk from the ready queue is assigned to the builder processes for processing. The supervisor process does not allow ready queue to grow indefinitely, leaving only the specified number of chunks in the queue and deleting old chunks. For deleting chunk, the supervisor process send all reader processes chunk drop command. The removal of chunks indicates a problem in the slice building system, although this is a valid operating mode of the system (for example, during commissioning). The queue size should be chosen to prevent full chunks from being lost when there are temporarily insufficient builder processes. Additionally, it should ensure incomplete chunks do not occur if one reader process receives a full chunk while another lacks space for it.

In response from the builder process, the supervisor process sends it the run ID and the chunk number to process. The supervisor process monitors the network load on builders' computers based on the timestamps of the beginning and end of chunk transfer from the reader processes to the builder process. This accounting allows one to run multiple builder processes on the same builder computer. In a situation where there are several applicants for processing one chunk, the supervisor's process can make a choice which builder process will spend less time reading chunk from all the reader processes, based on the workload of network at the time of the decision.

The supervisor process is controlled by a run control system, for which the entire slice building system looks like a single whole and is controlled through the supervisor process.  The main commands received from the run control system include starting/stopping the run, pausing/resuming the run.

The builder process reads data belonging to one chunk from all reader processes, creates a complete chunk, writes it to the data storage and makes the corresponding entry in the database.
Builder processes run on builder computers, which will be located near the data storage. This determines the location of the builder computers in the NICA data center.

At startup, the builder process registers with supervisor. The number of builder processes can change dynamically: new builder processes can be registered at any time, and some builder processes may terminate naturally (when the builder process completes processing one chunk and does not request a new one) or as a result of an execution error. A problem with a single builder process will not be a problem for slice building systems as a whole if there is some reserve of builder processes, but it may lead to loss of chunk data. When the builder process is ready to process the chunk, it requests the chunk from the supervisor process. As a result, the builder process will receive the run ID and the chunk number for processing. After that, the builder process connects to all reader processes and starts receiving chunk data. Upon completion of receiving the data, the builder process closes  the connections to the reader processes, and the actual building of the full chunk begins. The assembled chunk is stored in an intermediate data storage, and the corresponding entry is entered into the database. In parallel with saving the chunk data, the build process sends the specified number of slices to the main distributor's server. The builder process informs the supervisor process about the start of data transfer from the reader processes, about the completion of the transfer, and about writing the assembled chunk to an intermediate data storage.

\section{Summary and conclusions}
The data acquisition system for the SPD experiment at NICA has been conceptually designed to respond the stringent requirements imposed by the experimental setup and the anticipated data high rate. 
As a hardware trigger system under the specific high collision rate hardly can be realized, the free-running (triggerless) data acquisition system is chosen.  This provides continuous data recording approach which accepts the expected data flux.   

The DAQ system is structured into three main subsystems: the Readout Chain, the Time Synchronization System, and the Slice Building System. The system is being developed to ensure high reliability and scalability. The TSS, based on White Rabbit technology, ensures sub-nanosecond synchronization accuracy across all subdetectors, enabling consistent and precise timestamping. The Readout Chain guarantees efficient data collection and transmission, while the Slice Building System organizes the acquired data into time-based segments for subsequent analysis and storage.

The scalability of the system is demonstrated by its capacity to accommodate the increasing number of detector channels as the experiment progresses through its phases. For the initial stage, the system is designed to handle approximately 180000 channels, scaling up to 600000 channels in the full implementation. With the expected data flux of up to 20 GByte/s, the system's architecture is optimized to ensure minimal overhead while maintaining the integrity of transmitted data.

The data format has been designed to balance the inclusion of detailed event information with efficient use of bandwidth and processing resources. 

The proposed DAQ system provides a robust and scalable solution tailored to the specific challenges of the SPD experiment. Its conceptual design ensures high throughput, precise synchronization, and adaptability, laying a strong foundation for addressing the demanding requirements of SPD. The proposed DAQ system can also be efficiency used in other experimental setups operating under similar conditions.
\section{Acknowledgments}
The authors are grateful to our colleagues from SPD collaboration for helpful discussions, especially to Dr. Danila Oleynik for his expert advices.

This work was partially supported by Ministry of Science and Higher Education of the Russian Federation, state assignment for fundamental research (code FSEG-2025-0009) in the part of the TSS subsystem for time synchronization. Also the research was carried out within the framework of the state assignment of the Ministry of Science and Higher Education of the Russian Federation (project No. FSWM-2025-0023) in the part of study of the L2 concentrator for readout chain.

\bibliographystyle{unsrt}
\bibliography{references}

\end{document}